\documentclass{iopart}

\usepackage{iopams}
\usepackage{epsfig}
\usepackage{floatflt}

\newcommand{\newc}{\newcommand}
\newc{\ETJ}{E^{{\rm Jet}}_T}
\newc{\xgo}{x_\gamma^{\rm OBS}}
\def\ptmin{\hat{p}_T^{\rm min}}

\begin{document}

\jl{4}

\title{Soft Particle Emission Accompanying Dijet Photoproduction}
\author{Jonathan M. Butterworth$^1$, Valery A. Khoze$^2$ and Wolfgang Ochs$^3$\\}
\address{
$^1$ Department of Physics and Astronomy, University College London\\
$^2$ INFN - Laboratori Nazionali di Frascati\\
$^3$ Max-Planck-Institut f\"ur Physik, Munich
}

\begin{abstract}
The intensity of the soft bremsstrahlung 
depends on the colour charges of the hard- scattered partons. 
This intensity is larger in
dijet production with resolved photons by the factor $C_A/C_F=9/4$
in comparison with direct photoproduction for particular
configurations.
 It is investigated
how these effects are reflected in Monte Carlo models. 
\end{abstract}


\noindent

The idea that the shape of inclusive hadron spectra can be derived from
the perturbatively calculated parton distributions 
(Local Parton Hadron Duality  \cite{adkt1})
has been quite successful in general. There are some specific  predictions for 
the very soft particles. 
 Because of
their long wavelength the soft gluons ``probe'' only the colour factors 
of the primary hard partons (the ``colour antennae patterns'' \cite{adkt1})
 and then their production properties
can be derived directly from the lowest order perturbative diagrams.
Recently the case of soft particles has been investigated 
in more detail  \cite{klo1}.
In the simplest case of $e^+e^-\to q \overline q$ the density of the
soft gluons is predicted to be practically energy independent. 
Remarkably, it was found
that the soft hadrons indeed follow this expectation 
suggesting the relevance of the duality picture even for
low momentum particles ($p <\sim 300$ MeV).

This success could be accidental and may reflect purely hadronic
properties.  It is therefore important to test more specifically the
perturbative origin of the soft phenomena. A direct consequence of the
dominance of the lowest order diagrams is the proportionality of the
soft particle density to the primary parton colour factor which is
$C_F=\frac{4}{3}$ and $C_A=3$ for a $q\overline q$ and $gg$ antenna
respectively. Then low momentum particles are produced with the
relative intensity $\frac{9}{4}$. It is important to recall that the
total event multiplicities approach this ratio rather slowly with
energy. As a $gg$ system is difficult to prepare one has to find an
equivalent parton antenna with the appropriate effective colour
factors. Such a realization is provided by, for example, dijet
photoproduction.

\section{Tests with Photoproduction of Dijets} 
One can distinguish in the leading order QCD approach the direct and
the resolved processes.  In the first case the photon participates
directly in the hard scattering subprocess by photon-gluon-fusion or
QCD-Compton scattering and transfers a large fraction
($x_\gamma\sim1$) of its primary energy to the secondary jets. In the
second case, the hard scattering subprocess involves the partons
($q,\bar q$ and $g$) in the photon and in the proton and the energy
fraction $x_\gamma < 1$.  At HERA the dominant direct and resolved
processes correspond to quark and gluon exchange respectively and the
expected distributions in the dijet $cms$ scattering angle $\Theta_s$
have been clearly observed at HERA \cite{dijzeus}.

Particularly simple results arise for the soft radiation
perpendicular to the scattering plane \cite{klo1}. In this case complications 
with the cut-off
in the transverse momentum $k_\perp\geq Q_0$ disappear and all formulae
depend only on the angles between the hard partons.
In the simplest case of a $q\overline q$
dipole the density of gluons with momentum $p$
perpendicular to $q\overline q$ is  
%
\begin{equation}
\frac{dN_{q\bar q}}{d\Omega dp} 
   = \frac{\alpha_s }{(2\pi)^2p}\;{2C_F} (1-\cos\Theta_{q \overline q}).
\label{dipole}
\end{equation}
The soft radiation in more complicated hard processes involving gluons, 
relevant in the present discussion, can
be obtained from appropriate superpositions of 
elementary dipoles.

We consider the ratio $R_\perp^i$ of the perpendicular radiation 
in the process $i$ and in a standard $q\overline q$ process with
$\Theta_{q \overline q}=\pi$. 
With this normalization one finds for the direct processes 
a) $\gamma g \to q \overline q$ and b) $ \gamma q \to q g$ 
\begin{equation}
R_\perp^a\ = 1,    
\qquad
R_\perp^b\ =\  \frac{N_C}{4C_F}\left[3-\cos \Theta_s-\frac{1}{N_C^2}
   (1+\cos \Theta_s)\right].\label{rpb}
\end{equation}
In both cases $R_\perp^i\to 1$ for scattering angle 
$\Theta_s\to 0$, i.e. in this limit both processes
behave like  $q\overline q$ antennae. In the QCD-Compton
process b) the  $q\overline q$ antenna changes into a $gg$ type antenna
for $\Theta_s\to \pi$.

In case of the dominant resolved processes the results in Eqs. (58-61)
were  derived \cite{klo1} 
for small scattering angles. There is only a weak dependence on 
this angle
and for $\Theta_s\to 0$, 
where the gluon exchange dominates, they approach
\begin{equation}
R_\perp^i\ = \ C_A/C_F\ = \  2.25\; . \label{rpi}
\end{equation}
 Using the known full expressions for the soft gluon radiation patterns
\cite{dktemw} we found the approximate results for the leading processes 
 $ g g \to g g $ and $ q g \to q g$ to be correct within $\sim 10\%$ for
$\Theta_s<\pi/2$;
within this accuracy one can safely neglect the contribution from the
process  $g g\to q \overline q$ (see Eqs. (A7-A9) in [4c]). 
In the table we show the results for the ratios $R_\perp$ in the case when the
final partons are not identified, i.e. after symmetrization as in Eq. (53)
of \cite{klo1}. One can see that the result (\ref{rpi})  
remains
approximately valid at arbitrary scattering angles
within $\sim 20\%$.
%
\begin{center}
\begin{table}[b]
  \begin{tabular}{lccccccccccc}
\hline
 $\cos\ \Theta_s$ & 1.0 & 0.9 & 0.8 & 0.7 & 0.6 & 0.5 & 0.4 & 0.3 & 0.2 & 0.1
& 0.0 \\ \hline
 $\gamma g\ \to\ q \overline q$ & 1.0 & 1.0 & 1.0 & 1.0 & 1.0 & 1.0 & 1.0 & 1.0
 & 1.0 & 1.0 & 1.0 \\
  $\gamma q\ \to\ q g$ & 1.0 & 1.16 & 1.29 & 1.39 & 1.46 & 1.52 & 1.56 & 
1.59 & 1.61 & 1.62 & 1.62 \\
$gg\ \to \ gg$ & 2.25 & 2.31 & 2.36 & 2.41 & 2.46 & 2.51 & 2.55 & 2.58 & 
2.61 & 2.62 &  2.62 \\
$qg \ \to \ qg$ & 2.25 & 2.23 & 2.19 & 2.15 & 2.10 & 2.04 & 1.98 & 1.92 & 1.87 
& 1.84 & 1.83  \\ \hline
  \end{tabular}
\caption{Angular dependence of symmetrized cross section ratios $R_\perp$
for quark and leading gluon exchange processes}
 \label{tab1}
\end{table}
\end{center}

\section{Monte Carlo Studies} 

We have made a `Monte Carlo measurement' of this ratio using the
HERWIG~\cite{HERWIG} program. HERWIG contains coherent QCD radiation
effects and a cluster hadronization model. However, since HERWIG
produces a complete simulation of events, realistic jet algorithms and
detector acceptance cuts can be applied. Thus, if HERWIG is consistent
with the analytical results, this is an indication that the
perturbative calculations are relevant for soft particles as well,
that the results are insensitive to detector acceptance effects, and
that the measurement is likely to be feasible.

HERWIG events with a hard scatter of $p_T > 6$~GeV were generated for
HERA beam conditions in the region $0.2 < y < 0.85, Q^2 < 1$~GeV$^2$,
where $y$ is the usual inelasticity variable and $Q^2$ is the
virtuality of the photon exchanged between the proton and the
positron. The $k_T$ jet finder was run in inclusive mode~\cite{kt} on
the hadronic final state, and two jets with $\ETJ \ge 6$, at least one
of which must have $\ETJ > 7$~GeV, were demanded, in the
pseudorapidity region $|\eta| < 2$. We have also imposed the cut
$|\eta_1 + \eta_2|/2 < 1$, constraining the boost of the dijet system
so to achieve a more uniform acceptance in jet scattering angle, as
described in~\cite{dijzeus}.  Together, these cuts correspond to a
region in which measurements have previously be made at HERA.  Only
$4.5$~pb$^{-1}$ of simulated data was used.

The particle $p_T$ spectrum down to 50~MeV for particles within a cone
of one unit in $\eta-\phi$ from the vector perpendicular to the dijet
system in the dijet centre of mass system is then obtained.

Based upon the variable $\xgo$~\cite{dijzeus}, the fraction of the
photon's momentum which enters into the dijet system, we then divide
the events into two samples - $\xgo > 0.75$ (`resolved') and $\xgo <
0.75$ (`direct'). The ratio of the $p_T$ spectra for resolved/direct
events is calculated, and is shown in the upper two figures for two
regions of scattering angle, large (left) and small (right) - solid
points.  The small scattering angle region is defined by $|\Delta\eta|
> 2$ and the large by $|\Delta\eta| < 2$. In terms of $\cos\Theta_s$,
these ranges correspond approximately to $0 < \cos\Theta_s < 0.76$ and
$0.76 < \cos\Theta_2 < 0.96$. The plots also show the predictions
above for these kinematic ranges. The line shows the prediction taking
the subprocess mixture as given by HERWIG. The band shows the
uncertainty in the prediction if absolutely no knowledge of the
partonic subprocess type, but a perfect separation of resolved and
direct type diagrams is assumed. In all cases, processes not included in
the table are neglected.

\begin{figure}[ht]
\begin{center}
\psfig{file=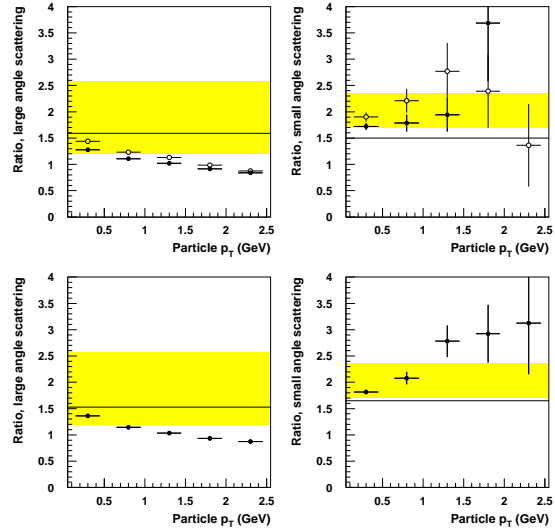,height=8.0cm}
\protect\caption{\it Ratio of the average track transverse momentum in high
$\xgo$ to low $\xgo$ events (i.e. `Resolved/Direct') for large
jet scattering angles (left) and for small jet scattering angles (right).}
\end{center}
\end{figure}

The HERWIG results approach the analytic predictions at low $p_T$.
Also shown (clear circles) is the result from HERWIG when multiparton
interactions are allowed~\cite{MI} with a $\ptmin$ for the hard
scatter of 2.5~GeV. Multiparton interactions are often appealed to as
a means of improving agreement between data and MC in the HERA
region. They raise the ratio slightly, moving it closer to the
prediction for high angle scattering, and further away for small angle
scattering.

Because the resolved cross section peaks strongly at low scattering
angles while the direct matrix element does not, the high $\xgo$
sample for small scattering angles in fact consists of 30\%
resolved-type diagrams. This is why the `prediction' lies outside the
band, since in constructing the band it was assumed that the
separation between resolved and direct LO diagrams was perfect. The
extra contamination from resolved acts against the enhancement in the
ratio coming from the angular dependence of the radiation. Clearly a
better way of distinguishing resolved and direct type diagrams is
required - either a new variable, or a harder cut on $\xgo$. We have
tried cutting at $\xgo = 0.9$, at the same moving the separation
between low and high angular regions to $|\Delta\eta| = 1.5
(\cos\Theta_s = 0.64)$. The results are shown in the lower plots.  The
prediction for low angle scattering is raised slightly and the
agreement with HERWIG is rather good. So it looks like for
these distributions the HERWIG results reproduce duality picture quite
closely.  Further improvements in these tests are undoubtedly possible.

The results presented reiterate that it is of great interest to
measure these distributions in HERA data. The direct comparison of the
low momentum particle production in direct and resolved processes
should yield the ratio of colour factors $C_A/C_F$. Its observation is
a crucial test of the significance (or otherwise) of multiparton
interaction models and, more fundamentally, of the perturbative
picture for soft particle production.

\end{document}